# Sparks of Quantum Advantage and Rapid Retraining in Machine Learning

William Troy[1*]

## Abstract

The advent of quantum computing holds the potential to revolutionize various fields by solving complex problems more efficiently than classical computers. Despite this promise, practical quantum advantage is hindered by current hardware limitations, notably the small number of qubits and high noise levels. In this study, we leverage adiabatic quantum computers to optimize Kolmogorov-Arnold Networks, a powerful neural network architecture for representing complex functions with minimal parameters. By modifying the network to use Bézier curves as the basis functions and formulating the optimization problem into a Quadratic Unconstrained Binary Optimization problem, we create a fixed-sized solution space, independent of the number of training samples. This strategy allows for the optimization of an entire neural network in a single training iteration in which, due to order of operations, a majority of the processing is done using a collapsed version of the training dataset. This inherently creates extremely fast training speeds, which are validated experimentally, compared to classical optimizers including Adam, Stochastic Gradient Descent, Adaptive Gradient, and simulated annealing. Additionally, we introduce a novel rapid retraining capability, enabling the network to be retrained with new data without reprocessing old samples, thus enhancing learning efficiency in dynamic environments. Experiments on retraining demonstrate a hundred times speed up using adiabatic quantum computing based optimization compared to that of the gradient descent based optimizers, with theoretical models allowing this speed up to be much larger! Our findings suggest that with further advancements in quantum hardware and algorithm optimization, quantum-optimized machine learning models could have broad applications across various domains, with initial focus on rapid retraining.

## 1. Introduction

The advent of quantum computing (QC) promises to revolutionize various fields by solving complex problems more efficiently than classical computers. This is possible as quantum computers leverage the principles of superposition and entanglement to perform computations that would be infeasible for classical computers, potentially offering exponential speedups for certain types of problems[1–3]. For instance, Shor's algorithm for factoring large numbers can theoretically break widely used cryptographic systems much faster than the best classical algorithms[4]. Despite this promise, the practical realization of quantum advantage is hindered by current limitations in quantum hardware, notably the relatively small number of qubits and high noise levels on modern quantum processors. These limitations make it challenging to solve large-scale problems and require highly optimized algorithms to make the most of the existing quantum resources[5]. To date, quantum advantage has only been demonstrated a couple of times, notably by Google's Sycamore processor and in Gaussian boson sampling experiments[6,7].

[1] Independent Researcher
[*] Corresponding Author: troywilliame@gmail.com

While these achievements mark significant milestones, quantum advantage in other domains remains difficult to achieve. A particularly promising area of application for QC is quantum machine learning (QML), which integrates quantum computing with machine learning techniques to leverage quantum speedups for training and inference tasks[8–10]. Platforms like TensorFlow and Qiskit have already started integrating quantum models, providing tools for researchers to explore QML applications[11,12]. Despite these advancements, Google recently highlighted a significant gap in the field at Google I/O 2024, stating that no one has yet demonstrated a clear quantum advantage for machine learning on classical data[11] This gap underscores the need for innovative approaches that can bridge this divide and showcase the practical benefits of quantum computing in machine learning.

In this study, we address this challenge by leveraging adiabatic quantum computing to optimize Bézier Kolmogorov-Arnold Networks (KANs). KANs represent a powerful neural network architecture known for their unique ability to represent complex functions with a relatively small number of parameters[14,15]. However, due to the complexities of the model the optimization of KANs is currently quite slow compared to multi-layered perceptions. By reformulating the optimization problem from that of a classical backpropagation algorithm to a Quadratic Unconstrained Binary Optimization (QUBO) problem we are able to create a fixed sized solution space that is independent of the number of training samples. Due to order of operations, a majority of the preprocessing of this QUBO problem can be simplified to use a collapsed version of the training dataset rather than the entire training dataset. Once preprocessing is done, D-Wave quantum annealers can be used as our adiabatic quantum computers. These quantum annealers are then able to find the optimal weights for this network, with the number of qubits and run time on the quantum annealers being independent of the number of training samples.

Our approach demonstrates sparks of quantum advantage when testing on small models which are able to fit within the currently limited number of qubits available. This demonstration is done by achieving faster training times compared to classical optimizers. These classical optimizers include one of the most popular optimization strategies in the Adam Optimizer[16–20] as well as Stochastic Gradient Descent[21] (SGD), Adaptive Gradient[22] (AdaGrad), and the classical parallel to quantum annealing in optimization by simulated annealing[23,24]. Moreover, we introduce a novel rapid retraining capability, enabled by the ability to retrain with new data without the need to reprocess old samples. This feature is particularly advantageous in dynamic environments where data is continuously evolving, such as real-time analytics, due to the fact that a quantum annealer can perform an entire optimization in tens of microseconds[25].

## 2. Results

We conducted a series of experiments to evaluate the performance of our Bézier-KANs optimized using different methods: backpropagation with the continuous variable gradient descent based optimizers via pytorch[26], QUBO with simulated annealing, and QUBO with quantum annealing. Our results cover five separate tasks which include two classification tasks with 1-layer KANs and three regression tasks using multi-layer KANs. Additionally, we explore the rapid retraining capability of our KAN formulation and its impact on training speed. Tasks chosen remain relatively simplistic due to the extremely limited number of quality qubits available in current hardware.

## 2.1. Classification Tasks

The classification tasks were chosen from common classification problem datasets available in SciPy[27] with one of them also being an example dataset in the pykan library[28]. For classification tasks we used 1-layer KANs of the shape [2,1], as this was the same structure used in the pykan classification examples. The training time and performance of the different optimization methods were evaluated based on accuracy, precision, recall, and F1-score, Figure 1. Despite the limited number of qubits used to create the discrete Bézier coefficients the quantum and simulated annealing achieve comparable performance to that of the continuous coefficient gradient descent based optimizers, Figures 1a and 1b. On top of this, the training time of the quantum method is significantly faster than all other tested methods. With the quantum method achieving training speeds over 12.5 times faster than the gradient descent based optimizers and almost 23 times faster than simulated annealing on the circle dataset!

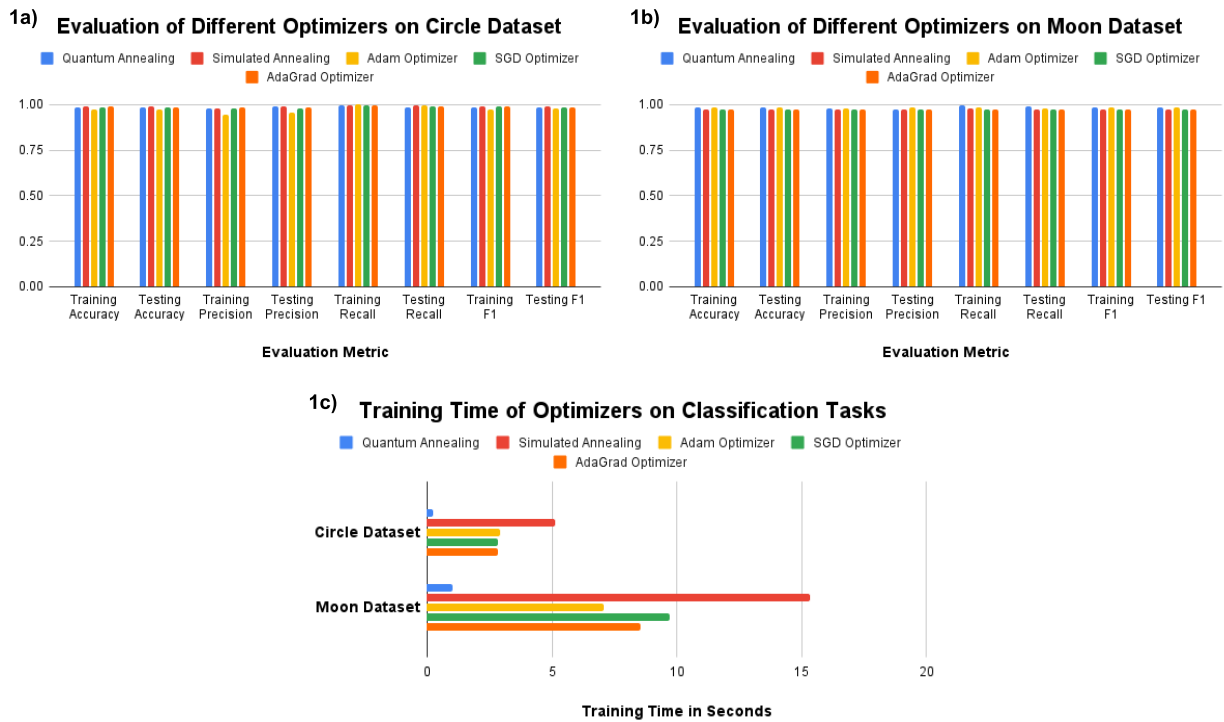

**Figure 1.** The training and testing metrics of Bézier-KANs optimized using classical optimizers and quantum annealing on (a) the circle dataset and (b) the moon dataset. (c) The respective training times for these different optimization scenarios.

## 2.2. Regression Tasks

For regression dataset 1, given by Eq. 20, and regression dataset 2, the spherical harmonic function, we evaluated the performance of the different optimization methods based on MSE and R-squared, Figure 2a-2d. Along with this in order to simulate an environment in which we may want to retrain our model several times we also performed rapid retraining. To do this we trained our initial model on both regression dataset 1 and 2. After which we then iteratively introduced new data to our respective datasets and retrained the models, paying attention to the time taken for each retraining and total time spent training the model, Figure 2e and 2f.

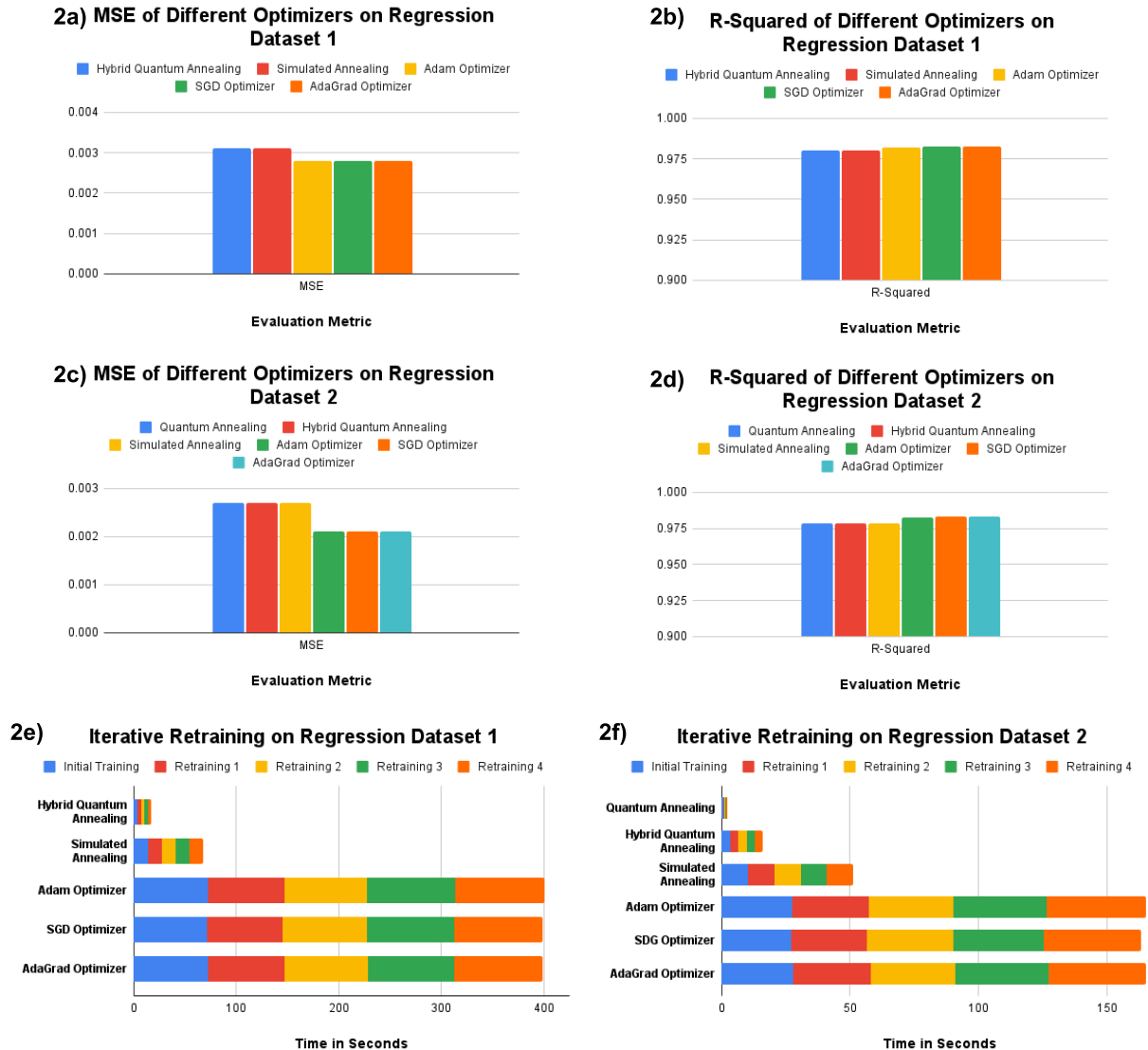

**Figure 2.** (a) MSE and (b) R-Squared values of Bézier-KANs trained on regression dataset 1 using classical optimizers and hybrid quantum annealing. (c) MSE and (d) R-Squared values of Bézier-KANs trained on regression dataset 2 using an Adam optimizer, simulated annealing, and pure quantum annealing. (e) Time taken for initial training and retraining using our different optimizers on regression dataset 1 and (f) on regression dataset 2. The initial training had 1,000,000 data points while each retraining added 100,000 new data points.

It should be mentioned that the hybrid quantum annealer used here finds solutions in around three seconds, as opposed to the tens of microseconds of a pure quantum annealer. It was necessary to use a hybrid quantum annealer for some of our regression tasks as they use a greater number of variables and hybrid quantum annealers can more consistently solve for a larger number of variables than current pure quantum annealers can. Despite this large time increase introduced to the quantum solutions by using a hybrid annealer, our hybrid method is able to perform all four rounds of iterative retraining in an eighth of the time taken to perform a single round of training by the gradient descent based optimizers for regression dataset 1! On top of this even the simulated quantum annealer dramatically out performs the gradient descent based optimizers on these tasks, due to not having to reprocess already seen samples. This is even taken a step further when looking at dataset 2. As the pure quantum annealer is able to perform the initial training 32 times faster than the gradient descent based optimizers and perform retraining in just 1% of the time taken by the gradient descent based optimizers!

Regression dataset 3, given by Eq. 21, involved a more complex KAN with a larger number of binary variables, over 200. Due to this large number of binary variables current quantum annealers struggled to encode the task well and converge, while the simulated annealer managed to find a solution if allowed to run enough times. With this we varied the degree of one of the Bézier curves on the bottom layer from one to two, leaving the rest of the network untouched with the other bottom Bézier curve and top Bézier curve having degrees of two and one, respectively, and performed 50 hybrid optimizations for both of our modified networks, Figure 3.

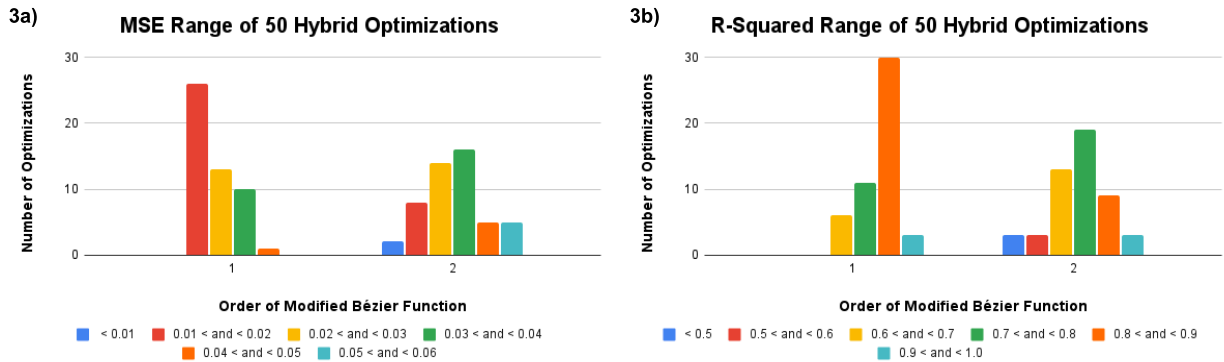

**Figure 3.** Distribution of (a) MSE and (b) R-Squared values for 100 hybrid optimizations of a Bézier-KAN, where the degree of a Bézier curve in the first layer was changed from 1 to 2.

From this it can be seen that against what one would expect if we increase the flexibility of the model to learn by increasing the degrees it has access to it actually can negatively affect the model by increasing the chance of finding a less optimal solution. Although this is unexpected it is attributed to the current state of quantum annealers in which the number of variables is limited while both noise and chain breaks are very prevalent. For comparison, simulated annealing was able to find a solution with a MSE value of 0.0041 and R-Squared of 0.963 for the Bézier curve being modified having a degree of one.

# 3. Discussion

Despite the limitations in the number and quality of current qubits, our work demonstrates that it is already possible to create machine learning frameworks with training speeds that outperform current classical methods in specific scenarios while maintaining comparable performance. This advantage is particularly evident when retraining networks with additional data. With pure quantum annealing, hybrid quantum annealing, and simulated annealing all performing retraining faster than the gradient descent based optimizers, with pure quantum annealing having the fastest retraining times. This success is achieved by creating fixed-sized solution spaces with states that can be saved and rapidly solved using an annealer. We anticipate further improvements in rapid retraining speed as advancements in quantum hardware allow pure quantum annealers to accurately solve more complex problems.

In theory we should be seeing even larger speedups. This is because all that is needed for our quantum optimization is a forward pass of the network before quantum annealing. Given that theoretical annealing runs in a fraction of a second regardless of the problem size, the main time component of the quantum optimization framework is the singular forward pass. Comparing this to classical back propagation algorithms that require both a forward and backward pass for each training epoch, it is easy to see that we should be seeing much greater experimental speedups. As our experiments require 50 to 150 epochs of classical training, even without including the time complexity of gradient calculations in backwards passes, one would expect to see speedups on the scale of the number of training epochs.

We attribute this discrepancy between theoretical speedups and what we see experimentally to code optimization. Furthermore, we believe there is a possibility for training speedups to go past this mark. As during the preprocessing step we can group all of our samples together and process entire datasets from a list of floating point numbers collapsed from the original training inputs, drastically reducing the amount of calculations done in the forward pass. Even in the current state of the code used to process this algorithm we are seeing emergences of this theoretical speedup going beyond that of a single iteration. This is demonstrated in regression task 2, where the entire processing for quantum optimization, 0.49 seconds, was faster than a singular iteration of any of the tested gradient based optimizers, 0.54 seconds.

There is also significant room for improvement of this algorithm with additional research focused on optimizing the quantum annealing process, reducing the number of auxiliary variables, problem decomposition, implementing similar approaches on gate-based quantum computers[29], and exploring the continuous optimization space, which is known to be simpler than that of the discrete space[30]. Perhaps the most promising area of all for advancing this methodology will be all-to-all connected quantum annealers[31,32]. All-to-all connectivity in quantum annealers promises to drastically reduce, if not eliminate, the need for auxiliary variables when solving HUBO problems[33]. This would theoretically eliminate the scalability issue of our algorithm caused by the increasing need for auxiliary variables as the number of layers in a network increases, allowing for a one-to-one qubit to variable representation of the entire system.

With these things in mind, we foresee this as just the first of a new class of quantum-optimizable machine learning models that create fixed-sized solution spaces. Exploring alternative KAN formulations using other spline functions, Taylor series expansions of spline basis functions, other objective functions, or entirely different machine learning models could lead to the development of more models that fit within this class.

As potential applications of this model type span various fields, particularly where rapid retraining can have a large effect, including finance, healthcare, environmental monitoring, and cybersecurity there is a large incentive to expand on and optimize our proposed model type. Along with this, although we already demonstrate experimental speedups in the training time of our model compared to classical optimization, continued advancements in quantum computing will naturally enhance the performance of our model. Allowing for a potential future in which complex machine learning networks could theoretically be trained and retrained on quantum computers at speeds dramatically faster than their classical counterparts are capable of.

## 4. Methods

KANs are a neural network architecture designed to represent complex functions with a minimal number of parameters. In order to achieve this representation, KANs build networks that are composed of basis functions. The original KAN representation used B-spline basis functions summed with weighted activation functions to manipulate edges of a network. However, it has also been shown that many other types of basis functions including Fourier-splines[34] and wavelets[15] can also work to create KANs. While we are able to use our end formulation to optimize one layer KANs of many different basis functions including B-splines, fourier-splines, and gaussian radial basis function splines, supplementary material S1, these spline types are not able to fit into the QUBO framework when stacked. This is because they are composed of either non-differentiable base functions when fully expanded or functions whose variables are not able to be expressed in linear and quadratic terms alone. While this is not a problem for traditional optimization methods, as we will see, it becomes a significant issue when optimizing a multi-layered KAN using QUBO methods.

### 4.1. Quadratic Unconstrained Binary Optimization (QUBO) Problem

QUBO is a mathematical formulation used to represent optimization problems where the objective is to find the minimum of energy a quadratic polynomial over binary variables. These problem setups represent energy functions that are extremely well-suited for quantum annealers, such as those developed by D-Wave, because they can exploit quantum parallelism, quantum tunneling, and quantum entanglement to search for the global minimum more efficiently than classical methods. In general for quantum annealers this energy function can be written as:

$$f(\vec{q}) = \vec{q}^T Q \vec{q} \qquad (1)$$

where $Q$ is an upper matrix, $\vec{q} = (q_1, ..., q_N)^T$, and $q_i$ represents binary variables. Given that $q_i \in [0, 1]$, $q_i^2 = q_i$, which leads us to the QUBO formulation of this energy function:

$$f(\vec{q}) = \sum_{i=1}^{N} Q_{i,i} q_i + \sum_{i<j}^{N} Q_{i,j} q_i q_j \tag{2}$$

where $Q_{i,i}$ and $Q_{i,j}$ denote the diagonal and off-diagonal terms, respectively, of a QUBO matrix[35].

These binary variables in our case represent the control points in our basis functions that make up the KANs. These continuous control points can be made into binary variables using multiple forms of binary discretization, commonly in the form radix 2:

$$x_i \approx \sum_{l=-m}^{m} 2^l q_{i,l} \tag{3}$$

where $x_i$ is any positive real integer and $2m + 1$ is the number of qubits used to represent $x$[36]. From this one can see that this can be expanded to include negative numbers by creating a $q^+$ and $q^-$ to represent positive and negative values respectively:

$$x_i \approx \sum_{l=-m}^{m} 2^l q_{i,l}^+ - 2^l q_{i,l}^- \tag{4}$$

While QUBO problems are ideal for quantum annealing, the formatting of a KAN optimization problem into one presents several unique challenges. The first difficulty is that variables of interest are required to be able to be expressed in only linear or quadratic terms. The second difficulty lies in the need to fully expand the basis functions of a KAN into their component terms. This expansion is crucial as the entire KAN can be thought of as one large equation with each layer feeding into the next. This leads to a problem when functions used to make a layer are discontinuous and reliant on one knowing the output of the previous layer. Unlike traditional neural network optimization, where you know the current output of a layer, as weights are iteratively adjusted to minimize the loss function via gradient descent, QUBO does not initialize the weights before solving. As QUBO solves for all weights at the same time, on what can be thought of as a black box, this approach eliminates the iterative modification of variables which is so time consuming in backpropagation. Although, this QUBO formulation brings in the extra requirements of variables being expressed in linear or quadratic terms and of continuous differentiability across all functions involved.

**4.2. Bézier Curves as a Solution**

To address these issues presented by QUBO formulation, we utilize Bézier curves as the basis functions for our KANs. Bézier curves are continuously differentiable in their expanded form and allow variables to be expressed in terms of just coefficients and control points, making

them suitable for multi-layer optimization using QUBO methods. With a Bézier curve of degree $n$ is given by:

$$B(t) = \sum_{i=0}^{n} \binom{n}{i} (1-t)^{n-i} t^i P_i \qquad (5)$$

where $P_i$ are the control points and $t \in [0, 1]$. To formulate this into a QUBO format the continuous $P$ variables are then discretized using variations of radix 2 encoding. This makes the theoretical number of logical qubits needed to solve the problem directly related to the number of control points in a given problem and the number of binary variables used to represent each control point.

### 4.3. Formulation of The QUBO Objective Function

In order to create a QUBO optimization function one must define an objective function ($OBJ$) to be optimized. In our case we used the mean squared error (MSE) as our machine learning objective. MSE was chosen as unlike many other traditional machine learning objective functions, it is fully continuous. Making our objective take the form of:

$$OBJ = \frac{1}{n} \sum_{i=1}^{n} \left(y_i - \widehat{y_i}\right)^2 \qquad (6)$$

where $n$ in the number of training samples, $y_i$ is the desired output at sample $i$, and $\widehat{y_i}$ is the network output. This objective function is a perfectly valid one and is commonly used in machine learning, however, if optimized to its fullest extent it can easily lead to overfitting of the training data. Since we cannot check for overfitting between training iterations, as we can classically, we re-formulate the objective function so that it pays attention to a validation dataset on its own:

$$OBJ = \frac{1}{n_t} \sum_{t=1}^{n_t} \left(y_t - \widehat{y_t}\right)^2 + \frac{\lambda}{n_v} \sum_{v=1}^{n_v} \left(y_v - \widehat{y_v}\right)^2 \qquad (7)$$

with $n_t$ and $n_v$ denoting the number of training and validation samples, respectively, and $\lambda$ representing the validation dataset weighting.

It should also be noted that while quantum noise is generally regarded as a negative, in the case of overfitting it can sometimes be viewed as a positive. This is because due to quantum noise, in most cases annealers will be run many times before landing on the most optimal solution. As such if the most optimal solution, with regards to the training dataset, is deemed as undesirable due to overfitting, or other metrics, one may be able to find a better solution for their specific metric of choice in the other solutions presented by the annealer.

### 4.4. Auxiliary Variables and HUBO Formulation

From this, one can see how this may lead to variable interactions beyond the quadratic order, the maximum order of which a QUBO formulation can solve for, creating a Higher Order Unconstrained Binary Optimization (HUBO) problem. This can be demonstrated in the theoretical case where we are optimizing a 2-layer Bézier-KAN of shape [1,1,1]. Using the syntax $B_{i,\ell}$ for our Bézier curves where $i$ is the index in the layer and $\ell$ is layer; we can define the entirety of the KAN as:

$$z = B_{1,2}(B_{1,1}) \tag{8}$$

If we then say all of the Bézier curves in this KAN are of order 1 we can define each Bézier curve as:

$$B_{i,\ell}(t) = (1 - t) P_{0,i,\ell} + t P_{1,i,\ell}, \quad 0 \leq t \leq 1 \tag{9}$$

If we then discretize the continuous control points using only the integer representation of radix 2 with two qubits, $q$, per control point we then end up with Eq. 10 for a given Bézier curve:

$$B_{i,\ell}(t) = (1 - t)\left(q_{0,0,i,\ell} + 2q_{1,0,i,\ell}\right) + t\left(q_{0,1,i,\ell} + 2q_{1,1,i,\ell}\right), \quad 0 \leq t \leq 1 \tag{10}$$

Substituting Eq 10 into Eq. 8 we get:

$$z = \left(1 - (1 - t)\left(q_{0,0,0,0} + 2q_{1,0,0,0}\right) + t\left(q_{0,1,0,0} + 2q_{1,1,0,0}\right)\right)\left(q_{0,0,0,1} + 2q_{1,0,0,1}\right) \\ + (1 - t)\left(q_{0,0,0,0} + 2q_{1,0,0,0}\right) + t\left(q_{0,1,0,0} + 2q_{1,1,0,0}\right)\left(q_{0,1,0,1} + 2q_{1,1,0,1}\right) \tag{11}$$

Which when multiplied out goes to:

$$\begin{aligned}
z = &\, q_{0,0,0,1} + 2q_{1,0,0,1} \\
&- q_{0,0,0,1}q_{0,0,0,0} - 2q_{0,0,0,1}q_{1,0,0,0} - 2q_{1,0,0,1}q_{0,0,0,0} - 4q_{1,0,0,1}q_{1,0,0,0} \\
&+ tq_{0,0,0,1}q_{0,0,0,0} + 2tq_{0,0,0,1}q_{1,0,0,0} + 2tq_{1,0,0,1}q_{0,0,0,0} + 4tq_{1,0,0,1}q_{1,0,0,0} \\
&- tq_{0,0,0,1}q_{0,1,0,0} - 2tq_{0,0,0,1}q_{1,1,0,0} - 2tq_{1,0,0,1}q_{0,1,0,0} - 4tq_{1,0,0,1}q_{1,1,0,0} \\
&+ q_{0,1,0,1}q_{0,0,0,0} + 2q_{0,1,0,1}q_{1,0,0,0} + 2q_{1,1,0,1}q_{0,0,0,0} + 4q_{1,1,0,1}q_{1,0,0,0} \\
&- tq_{0,1,0,1}q_{0,0,0,0} - 2tq_{0,1,0,1}q_{1,0,0,0} - 2tq_{1,1,0,1}q_{0,0,0,0} - 4tq_{1,1,0,1}q_{1,0,0,0} \\
&+ tq_{0,1,0,1}q_{0,1,0,0} + 2tq_{0,1,0,1}q_{1,1,0,0} + 2tq_{1,1,0,1}q_{0,1,0,0} + 4tq_{1,1,0,1}q_{1,1,0,0}
\end{aligned} \tag{12}$$

Now considering our objective function is MSE as defined in Eq. 6 we see that we must square Eq. 12. As this will quickly blow up into a much larger equation we will simplify it for this example and just take the last two terms in Eq. 12 and square them to see that the problem quickly becomes a higher order than quadratic with more than two $q$ variables being multiplied together. Keeping in mind a binary variable squared is equal to the original binary variable:

$$\left(2tq_{1,1,0,1}q_{0,1,0,0} + 4tq_{1,1,0,1}q_{1,1,0,0}\right)^2 =$$
$$4t^2 q_{1,1,0,1}q_{0,1,0,0} + 16t^2 q_{1,1,0,1}q_{0,1,0,0}q_{1,1,0,0} + 8t^2 q_{1,1,0,1}q_{1,1,0,0} \tag{13}$$

Even in this simplistic example where we only have three unique binary variables in Eq. 13, since everything is squared by the MSE objective function, variables become intertwined. In which auxiliary variables become necessary to bring the problem back to quadratic. This problem only becomes more pronounced when using higher-order Bézier curves within the KAN. As everything being input into each Bézier curve is raised to the degree of that curve, providing more opportunities to create a higher-order problem.

While this may seem like a major problem, one can easily get around it by introducing auxiliary variables with a singular variable being used to represent 2 or more others. An auxiliary variable that represents 2 other variables can be given by:

$$V_{AUX} = P_1 P_2 \tag{14}$$

With this relation being enforced by adding penalty terms the the QUBO problem in the form of:

$$Penalty = w\left(P_1 P_2 - 2V_{AUX}P_1 - 2V_{AUX}P_2 + 3V_{AUX}\right) \tag{15}$$

where $w$ is the penalty coefficient [37], 15-25 times the max coefficient value in our QUBO matrix in this study.

This makes the total number of logical qubits needed to optimize a network equal to the number of control points multiplied by the number of binary variables used to represent each control point plus the number of auxiliary variables:

$$num_{qubits} = num_{aux\_vars} + \sum_{i=1}^{num_{control\_vars}} num_{binary\_vars,i} \tag{16}$$

### 4.5. Collapsing The Inputs

As can be seen in Eqs. 12 and 13, once auxiliary variables are substituted in where necessary, the entire problem can be defined as a system of subsystems composed of 1-2 unknown variables multiplied by known variables, hence the definition of quadratic in a QUBO problem. Using this knowledge we can once again simplify the system to a combination of additions given by Eqs. 17 and 18:

$$subsystem = constant * [array] * unknowns \tag{17}$$

$$OBJ = subsystem_1 + subsystem_2 + subsystem_3 + ..... \tag{18}$$

Given that an array multiplied by a constant and then summed is equivalent to the array summed and then multiplied by a constant:

$$sum(constant * [array]) = constant * sum([array]) \tag{19}$$

We can simply precompute the possible arrays, which are just combinations of input and output arrays multiplied together, and collapse each of these possible combinations to singular floating point values. These collapsed values can then be put through the forward pass instead of putting entire arrays of inputs through them. This is then further sped up in practice if we consider that a majority of the known parts of the subsystem are reused throughout the entire system and that one can use memoization in order to not repeat these computations.

### 4.6. Rapid Retraining Capability

Given an initial dataset the expanded equation that represents the KAN is created and all available data points are put through a forward pass, after which objective function 1 ($OBJ_1$), is saved and a quantum annealer is used to solve for the weights given the input dataset(s). When new data becomes available, the expanded equation that represents the KAN and $OBJ_1$ is reloaded. The new data is put through a forward pass and the results of this forward pass are added to $OBJ_1$, creating $OBJ_2$. Once this is done the new $OBJ$, $OBJ_2$, is sent to a quantum annealer for optimization. With each trial of the optimizer taking tens of microseconds this creates an extremely fast retraining strategy without needing to reprocess already seen samples. Additionally this allows for the removal of samples if desired by simply putting then through a forward pass and then subtracting them from the $OBJ$.

### 4.7. Experimental Setup

We conducted experiments using two classification datasets and three regression datasets, each with varying sizes and complexities which could be solved given the limited number of available qubits. For classification tasks these datasets consist of the circle and moon datasets given by the popular SciPy package, each with 100,000 training and 10,000 testing data points. Regression tasks were selected from the original KAN paper, with regression dataset 1 given by Eq. 20, dataset 2 is given by the spherical harmonic function with m and n set to 0 and 1, respectively, and dataset 3 given by Eq. 21.

$$z = \frac{3x}{exp(y)+exp(-y)} \tag{20}$$

$$z = 2\sqrt{1 + x^2 + y^2} \tag{21}$$

The dataset inputs were normalized from 0 to 1 so that inputs are within the preferable range of a Bézier curve. Gradient descent based optimization was done using pytorch with a modified version of the original KAN library[28]. For all test cases involving the gradient descent based optimizers, only results with the optimal learning rate and number of steps were reported in the main text. The optimal gradient descent based learning rates were found by iteratively testing different learning rates for each problem from 0.001 to 1.5, with the 'best' denoting the learning rate that converged on the most optimal solution first. From this, the optimal number of steps was found by stopping learning once the loss function of the best learning rate converged for multiple training iterations.

For annealing, the objective problem formulation was done in c++ called from python which then used the pyqubo library[38,39] to convert the $OBJ$ to a QUBO matrix. Simulated annealing was done using the D-Wave simulated annealing library while quantum annealing used the D-Wave Advantage2 prototype 2.3 as a pure quantum annealer and hybrid binary quadratic model version 2 as a hybrid quantum annealer. Although quantum annealing can be run in tens of microseconds, the large amount of quantum noise currently present in quantum systems requires hundreds to thousands of quantum annealing runs to be done per problem. This is reflected in our experimental timings and makes the time taken by quantum annealing go from tens of microseconds to around a tenth of a microsecond for our test cases. Finally, since all optimization methods tested are highly amenable to parallel computing, this factor was eliminated and all classical computations were done on a single thread of an Intel Xeon Scalable Platinum 8173M Processor processor running at a 2 GHz clock speed.

**Software Availability Statement:**

Code used in this study is available at https://github.com/wtroy2/Quantum-KAN

**Conflict of interest:**

The authors declare that they have no conflict of interest.

# Sparks of Quantum Advantage and Rapid Retraining in Machine Learning Supplementary

## S1:

Many types of functions work for a single-layer KAN network that is to be put in QUBO format that do not work for a multi-layered KAN in QUBO format. This is due to the fact that for a one layer network at the time of doing the forward pass we know the value of all inputs to each basis function. Implementations of the B-spline, Fourier-spline, and Gaussian-Spline that are quantum optimizable on a simulated annealer are shown in Figure S1.

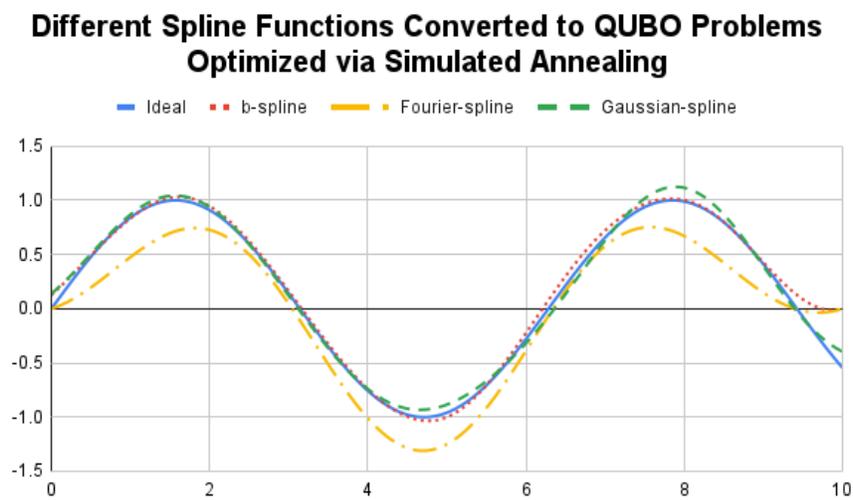

**Figure S1.** Different spline functions trained to fit to $y = sin(x)$ by creating an MSE objective function converted to QUBO format and solved using simulated annealing.

This becomes a problem when we start to stack layers as the output of one layer becomes the input to another. When this happens points we are solving for are either expressed as conditionals or in terms of functions which do not allow them to be expressed as either linear or quadratic, both of which are unacceptable for QUBO.